\documentclass[journal]{IEEEtran}

\usepackage{amsmath,amsfonts}
\usepackage{graphicx}
\usepackage{url}
\usepackage{hyperref}
\usepackage{array}
\usepackage{tabularx}
\usepackage{multirow}
\usepackage{algorithm}
\usepackage{algorithmic}
\usepackage[caption=false,font=footnotesize]{subfig}

\hypersetup{
  colorlinks=true,
  linkcolor=blue,
  citecolor=blue,
  urlcolor=blue
}

\graphicspath{{Figure/}}

\renewcommand{\arraystretch}{1}

\AtBeginDocument{%
  \setlength{\abovedisplayskip}{4pt plus 1pt minus 1pt}
  \setlength{\belowdisplayskip}{4pt plus 1pt minus 1pt}
  \setlength{\abovedisplayshortskip}{4pt plus 1pt minus 1pt}
  \setlength{\belowdisplayshortskip}{4pt plus 1pt minus 1pt}
}
\begin{document}

\title{CVaR-Guided Decision-Focused Learning and Risk-Triggered Re-Optimization for Two-Stage Robust Microgrid Operation}

\author{Tingwei~Cao, and Yan~Xu,~\IEEEmembership{Senior Member,~IEEE}%
\thanks{Tingwei Cao and Yan Xu are with Center for Power Engineering, School of Electrical and Electronic Engineering, Nanyang Technological University, Singapore 639798 (e-mail: \href{mailto:tingwei004@e.ntu.edu.sg}{tingwei004@e.ntu.edu.sg}; \href{mailto:xuyan@ntu.edu.sg}{xuyan@ntu.edu.sg}).}}

\maketitle

\begin{abstract}
Microgrid operation is highly vulnerable to short-term load uncertainty, while conventional predict-then-optimize pipelines cannot fully align probabilistic forecasting quality with downstream robust scheduling performance. This paper proposes a CVaR-guided decision-focused learning and risk-triggered re-optimization framework for two-stage robust microgrid operation. A probabilistic load forecasting model first generates multi-quantile outputs, which are converted into prediction intervals to parameterize the load uncertainty set of the downstream two-stage robust optimization (TSRO) model. To improve forecasting reliability under difficult and high-risk operating conditions, a CVaR-guided forecasting objective is introduced to emphasize tail-sensitive samples. To further close the forecast-decision gap, a convex regularized surrogate TSRO model and a smooth regret loss are developed, enabling downstream operational feedback to be propagated to the forecasting model through KKT-based implicit differentiation. For online deployment, a risk-triggered re-optimization mechanism selectively re-solves the remaining-horizon TSRO only when the schedule mismatch becomes significant, avoiding unnecessary online computation. Case studies on modified IEEE 33-bus and 69-bus microgrids demonstrate superior probabilistic forecasting accuracy, operational economy, and tail-risk mitigation over benchmark methods, while preserving near-full-re-optimization performance with less than 0.5\% higher operating cost and up to 91\% lower daily solution time.
\end{abstract}

\begin{IEEEkeywords}
Microgrid operation, probabilistic forecasting, two-stage robust optimization, decision-focused learning, conditional value-at-risk.
\end{IEEEkeywords}

\section{Introduction}
\IEEEPARstart{M}{icrogrids} coordinate renewable generation, energy storage systems (ESS), flexible loads, and grid interaction, and are widely regarded as an effective solution for enhancing local flexibility, efficiency, and resilience \cite{lasseter2011smart,guerrero2010hierarchical}. Their operation, however, remains highly exposed to short-term uncertainty. Schedules that appear economical under deterministic forecasts may become costly or difficult to sustain once actual operating conditions depart from the anticipated trajectory \cite{yang2021robust,he2023day,qiu2020historical}.

Robust optimization has been widely adopted for microgrid scheduling under uncertainty because it explicitly hedges against unfavorable realizations while remaining tractable \cite{zhang2016robust,zhang2017robust,yang2021robust,he2023day}. Among existing formulations, two-stage robust optimization is especially attractive, since it combines anticipative scheduling with adaptive recourse actions following uncertainty realization and thus matches practical microgrid operation \cite{zhang2016robust,zhang2017robust}. Existing TSRO-based studies have shown that coordinated recourse through ESS, direct load control (DLC), distributed generation, and demand response can improve operational robustness. However, many robust microgrid models still rely on uncertainty sets constructed from hand-crafted bounds, heuristic budgets, or coarse statistical assumptions \cite{yang2021robust,qiu2020historical}, which may not adequately capture the temporal variation and structure of actual operating uncertainty.

Probabilistic load forecasting provides a more informative interface between data and decision-making by producing quantiles or prediction intervals rather than only point estimates \cite{khajeh2022applications}. It has gradually been incorporated into microgrid scheduling under uncertainty, showing that data-driven uncertainty information can improve scheduling relative to rule-based uncertainty modeling. Nevertheless, most existing approaches still follow a sequential predict-then-optimize pipeline: the forecasting model is trained against statistical criteria, and its outputs are then passed to the downstream TSRO model \cite{ramahatana2022more,wang2023quantifying,AGUILAR2024123548,YANG2024134058}. This pipeline is limited because improved statistical accuracy does not necessarily translate into better downstream decision quality \cite{wang2023quantifying,doi:10.1287/mnsc.2020.3922}. Conventional forecasting losses mainly emphasize average performance, whereas extreme load deviations may dominate the upper tail of realized operating loss under robust scheduling. Recent value-oriented, cost-oriented, and risk-aware forecasting studies in power systems also suggest that such asymmetric tail effects should be reflected in the learning objective \cite{10219063,10771620,9616458}.

This observation is closely related to the predict-then-optimize mismatch that has motivated recent developments in predict-and-optimize and decision-focused learning (DFL) \cite{doi:10.1287/mnsc.2020.3922,doi:10.1287/ijoo.2023.0088}. Rather than training predictive models solely for statistical accuracy, these frameworks seek to improve downstream decision quality directly. Related ideas have begun to appear in power system applications and energy management problems \cite{9617122,11122623,WANG2026127343,ALKHULAIFI2026130554}. Meanwhile, recent microgrid studies have moved toward tighter forecast-decision coupling, including risk-aware scheduling frameworks for day-ahead operation \cite{AGUILAR2024123548,YANG2024134058,sheng2026risk}. Even so, a clear gap remains for TSRO-based microgrid operation: probabilistic load forecasts must both define meaningful uncertainty sets for robust scheduling and support end-to-end forecast-decision coupling through the downstream TSRO model. This dual requirement is nontrivial, especially when the original TSRO model contains binary variables and non-smooth recourse structures.

A second issue arises in online deployment. Re-solving the robust optimization problem over the remaining horizon at every dispatch step improves adaptability, but also imposes substantial online computational cost \cite{das2020novel,hu2023economic}. Continuing to execute an outdated schedule without timely correction may, in contrast, degrade both economy and robustness. Practical microgrid operation therefore requires tighter forecast-decision coupling together with selective online re-optimization.

Motivated by the above considerations, this paper proposes a Conditional Value-at-Risk (CVaR)-guided decision-focused learning and risk-triggered re-optimization framework for two-stage robust microgrid operation. The main contributions are summarized as follows:
\begin{itemize}
\item[1)] A CVaR-guided probabilistic load forecasting method is proposed to improve performance on tail events, yielding prediction intervals that remain more reliable under difficult and high-risk operating conditions \cite{khajeh2022applications,CAO2024123751,10219063,10771620,9616458}.

\item[2)] An end-to-end decision-focused training framework is established by coupling probabilistic load forecasting with a convex regularized surrogate TSRO model, allowing downstream operational feedback to be backpropagated to the forecasting model through a smooth regret loss and KKT-based implicit differentiation \cite{doi:10.1287/mnsc.2020.3922,doi:10.1287/ijoo.2023.0088,9617122,11122623,WANG2026127343,ALKHULAIFI2026130554}.

\item[3)] A risk-triggered re-optimization mechanism is designed for online operation. It re-solves the remaining-horizon TSRO only when the schedule mismatch becomes significant, thereby maintaining economic performance close to full re-optimization at a much lower online computational cost \cite{he2023day,AGUILAR2024123548,YANG2024134058}.
\end{itemize}

The remainder of this paper is organized as follows. Section~\ref{sec:problem_formulation} presents the probabilistic load forecasting model and the TSRO-based microgrid operation formulation. Section~\ref{sec:methodology} describes the proposed methodology, including CVaR-guided forecasting, surrogate decision-focused training, implicit differentiation, and risk-triggered re-optimization. Section~\ref{sec:case_study} reports the case studies and comparative results. Section~\ref{sec:conclusion} concludes the paper.

\section{Problem Formulation}
\label{sec:problem_formulation}

\subsection{Probabilistic Load Forecasting}

Accurate characterization of short-term load uncertainty is fundamental to robust microgrid operation. Unlike deterministic forecasting, probabilistic load forecasting aims to estimate the conditional distribution of future load demand rather than only a single point prediction.

At decision time $t$, let $\mathbf{X}_{t}$ denote the available input features, including historical load observations and exogenous information such as time labels and calendar-related variables. Given $\mathbf{X}_{t}$, a probabilistic load forecasting model parameterized by $\theta$ generates multi-quantile forecasts over the prediction horizon $\mathcal{T}=\{t+1,\ldots,t+H\}$ with dispatch interval $\Delta t$. Let $\mathcal{Q}=\{\alpha_{1},\ldots,\alpha_{M}\}$ denote the set of quantile levels. The forecasting model is written as
\begin{equation}
\hat{\mathbf{Q}}_{t}
=
\mathcal{F}_{\theta}\!\left(\mathbf{X}_{t}\right)
=
\left\{
\hat{q}_{\tau}^{(\alpha)}
\right\}_{\tau\in\mathcal{T},\,\alpha\in\mathcal{Q}}
\label{eq:pf_model}
\end{equation}
where $\hat{q}_{\tau}^{(\alpha)}$ denotes the predicted $\alpha$-quantile of the load demand at future period $\tau$.

Based on the multi-quantile outputs in \eqref{eq:pf_model}, the predictive median is taken as the nominal load forecast, and the corresponding prediction interval (PI) at confidence level $\kappa\in(0,1)$ is defined as
\begin{equation}
\hat{P}_{\tau}^{\mathrm{L}}=\hat{q}_{\tau}^{(0.5)},
\quad
\underline{P}_{\tau}^{\mathrm{L}}=\hat{q}_{\tau}^{(1-\kappa)/2},
\quad
\overline{P}_{\tau}^{\mathrm{L}}=\hat{q}_{\tau}^{(1+\kappa)/2}
\label{eq:predictive_median}
\end{equation}
\begin{equation}
\hat{\mathcal{I}}_{\tau}^{(\kappa)}
=
\left[
\underline{P}_{\tau}^{\mathrm{L}},
\overline{P}_{\tau}^{\mathrm{L}}
\right]
\label{eq:PI_def}
\end{equation}

Equations \eqref{eq:predictive_median} and \eqref{eq:PI_def} characterize the future load uncertainty used in this paper. The predictive median defines the nominal load trajectory, whereas the PI specifies the corresponding uncertainty range. These quantities serve as the basis for constructing the load uncertainty set in the subsequent TSRO formulation.

\subsection{Two-Stage Robust Microgrid Operation With Load Uncertainty and DLC}

Consider a grid-connected microgrid with wind turbine (WT) generation, photovoltaic (PV) generation, energy storage systems, direct load control, and power exchange with the main grid at the point of common coupling (PCC). In this paper, WT and PV outputs are treated as known inputs over the dispatch horizon, whereas only the load demand is modeled as uncertain. To coordinate the pre-scheduled operation and the real-time correction under load uncertainty, the microgrid operation is formulated as a two-stage robust optimization problem \cite{zhang2016robust}. In the first stage, a pre-dispatch schedule over the current optimization horizon is determined for the ESS and the grid exchange. In the second stage, after the load trajectory is realized within the uncertainty set, recourse actions are taken to adjust the scheduled grid exchange and ESS powers, activate DLC, and curtail renewable power if necessary so that all operating constraints remain satisfied. As in standard TSRO formulations, the second-stage variables are modeled as wait-and-see recourse variables with respect to the realized load trajectory over the current optimization horizon.

Let $x$ denote the first-stage schedule vector and let $y(u)$ denote the second-stage recourse vector under uncertainty realization $u$. Here, the uncertainty realization $u$ is equivalently represented by the admissible load trajectory $\{\tilde{P}_{\tau}^{\mathrm{L}}\}_{\tau\in\mathcal{T}}$ within the set $\mathcal{U}_{\mathrm{PI}}(\kappa)$. The TSRO model is written in the compact form
\begin{equation}
\min_{x\in\mathcal{X}} \ \max_{u\in\mathcal{U}_{\mathrm{PI}}(\kappa)} \ \min_{y(u)\in\mathcal{Y}(x,u)} \ \Phi(x,y(u);u)
\label{eq:tsro_compact}
\end{equation}
where $\Phi(x,y(u);u)$ is the total operating cost over the dispatch horizon.

The load uncertainty set is constructed directly from the probabilistic load forecasting outputs in \eqref{eq:predictive_median} and \eqref{eq:PI_def} \cite{YANG2024134058}. Let $\tilde{P}_{\tau}^{\mathrm{L}}$ denote an admissible load realization at period $\tau$. The interval-based uncertainty set is defined as
\begin{equation}
\mathcal{U}_{\mathrm{PI}}(\kappa)=
\left\{
\tilde{P}_{\tau}^{\mathrm{L}}:
\underline{P}_{\tau}^{\mathrm{L}}
\le
\tilde{P}_{\tau}^{\mathrm{L}}
\le
\overline{P}_{\tau}^{\mathrm{L}},
\ \forall \tau\in\mathcal{T}
\right\}
\label{eq:U_PI}
\end{equation}

The first-stage decision vector is defined as
\begin{equation}
x=
\left\{
P_{\tau}^{k,\mathrm{sch}}
\right\}_{k\in\mathcal{K}_{r},\tau\in\mathcal{T}}
\label{eq:first_stage_vector}
\end{equation}
where $\mathcal{K}_{r}=\{\mathrm{buy},\mathrm{sell},\mathrm{ch},\mathrm{dis}\}$ denotes the set of scheduled grid-exchange and ESS variables.

To exclude physically meaningless schedules, the first-stage vector $x$ is required to be feasible under the forecasted base-case load trajectory $\hat{P}_{\tau}^{\mathrm{L}}$. Accordingly, the feasible set $\mathcal{X}$ is defined through the auxiliary nominal variables and nominal feasibility conditions in \eqref{eq:nominal_aux}--\eqref{eq:nominal_initial_energy}:
\begin{equation}
\bar{y}=
\left\{
E_{\tau}^{\mathrm{sch}},
P_{\tau}^{\mathrm{cur,WT,sch}},
P_{\tau}^{\mathrm{cur,PV,sch}}
\right\}_{\tau\in\mathcal{T}}
\label{eq:nominal_aux}
\end{equation}
\begin{equation}
\begin{aligned}
&P_{\tau}^{\mathrm{WT}}
-
P_{\tau}^{\mathrm{cur,WT,sch}}
+
P_{\tau}^{\mathrm{PV}}
-
P_{\tau}^{\mathrm{cur,PV,sch}}
\\
&\quad
+
P_{\tau}^{\mathrm{dis,sch}}
-
P_{\tau}^{\mathrm{ch,sch}}
+
P_{\tau}^{\mathrm{buy,sch}}
-
P_{\tau}^{\mathrm{sell,sch}}
=
\hat{P}_{\tau}^{\mathrm{L}}
\end{aligned}
\label{eq:nominal_power_balance}
\end{equation}
\begin{equation}
E_{\tau+1}^{\mathrm{sch}}
=
E_{\tau}^{\mathrm{sch}}
+
\eta^{\mathrm{ch}} P_{\tau}^{\mathrm{ch,sch}} \Delta t
-
\frac{1}{\eta^{\mathrm{dis}}} P_{\tau}^{\mathrm{dis,sch}} \Delta t
\label{eq:nominal_ess_dynamics}
\end{equation}
\begin{equation}
E_{t}^{\mathrm{sch}} = E_{t}^{0}
\label{eq:nominal_initial_energy}
\end{equation}

For compactness, the nominal network variables associated with the forecasted base-case schedule are not explicitly included in $\bar{y}$, but they are required to satisfy the nominal counterparts of the network feasibility constraints introduced later.

Furthermore, the scheduled variables in $x$ and $\bar{y}$ must inherently satisfy the nominal counterparts of the physical component capacities, network limits, and energy bounds detailed later in the operational constraints.

After the load trajectory is realized, the second-stage recourse vector is defined as
\begin{equation}
\begin{aligned}
y(u)=\,
&\left\{
\Delta P_{\tau}^{k,+},
\Delta P_{\tau}^{k,-}
\right\}_{k\in\mathcal{K}_{r},\tau\in\mathcal{T}}
\\
&\cup
\left\{
E_{\tau},
P_{\tau}^{\mathrm{dlc}},
P_{\tau}^{\mathrm{cur,WT}},
P_{\tau}^{\mathrm{cur,PV}}
\right\}_{\tau\in\mathcal{T}}
\end{aligned}
\label{eq:second_stage_vector}
\end{equation}
where $\Delta P_{\tau}^{k,+}$ and $\Delta P_{\tau}^{k,-}$ denote the upward and downward recourse adjustments with respect to the first-stage schedule. For compactness, the network recourse variables $\{P_{ij,\tau},Q_{ij,\tau},V_{i,\tau}\}$ are not explicitly included in \eqref{eq:second_stage_vector}, but are understood as part of the second-stage feasible set $\mathcal{Y}(x,u)$. For notational simplicity, the dependence of the second-stage variables on $u$ is omitted hereafter.

The realized operating powers and the corresponding operating cost are jointly determined by the first-stage schedule and the second-stage recourse. Specifically, for all $k\in\mathcal{K}_{r}$ and $\tau\in\mathcal{T}$,
\begin{equation}
P_{\tau}^{k}
=
P_{\tau}^{k,\mathrm{sch}}
+
\Delta P_{\tau}^{k,+}
-
\Delta P_{\tau}^{k,-}
\label{eq:link_compact}
\end{equation}
and the total operating cost in \eqref{eq:tsro_compact} is expressed as
\begin{equation}
\Phi(x,y;u)=
\sum_{\tau\in\mathcal{T}}
\left(
C_{\tau}^{\mathrm{DA}}(x)
+
C_{\tau}^{\mathrm{RT}}(y)
\right)
\label{eq:cost_function}
\end{equation}

\begin{equation}
\begin{aligned}
C_{\tau}^{\mathrm{DA}}(x)=\,
&c_{\tau}^{\mathrm{buy,DA}} P_{\tau}^{\mathrm{buy,sch}}
-
c_{\tau}^{\mathrm{sell,DA}} P_{\tau}^{\mathrm{sell,sch}}
\\
&+
c^{\mathrm{ch}} P_{\tau}^{\mathrm{ch,sch}}
+
c^{\mathrm{dis}} P_{\tau}^{\mathrm{dis,sch}}
\end{aligned}
\label{eq:dayahead_cost}
\end{equation}
\begin{equation}
\begin{aligned}
C_{\tau}^{\mathrm{RT}}(y)=\,
&\sum_{k\in\mathcal{K}_{r}}
\left(
\pi_{k}^{+}\Delta P_{\tau}^{k,+}
+
\pi_{k}^{-}\Delta P_{\tau}^{k,-}
\right)
\\
&+
c^{\mathrm{dlc}} P_{\tau}^{\mathrm{dlc}}
+
c^{\mathrm{cur}}
\left(
P_{\tau}^{\mathrm{cur,WT}}
+
P_{\tau}^{\mathrm{cur,PV}}
\right)
\end{aligned}
\label{eq:realtime_cost}
\end{equation}
where $C_{\tau}^{\mathrm{DA}}(x)$ and $C_{\tau}^{\mathrm{RT}}(y)$ denote the day-ahead scheduled cost and the real-time recourse cost at period $\tau$, respectively. Here, $c_{\tau}^{\mathrm{buy,DA}}$ and $c_{\tau}^{\mathrm{sell,DA}}$ denote the day-ahead settlement prices of scheduled grid purchase and sale, whereas $\pi_{k}^{+}$ and $\pi_{k}^{-}$ denote the real-time penalty coefficients associated with deviations from the first-stage schedule. Therefore, the first-stage schedule determines the base operating plan under day-ahead settlement, while the second-stage adjustment is penalized separately.

The second-stage recourse decisions are subject to the following operational constraints for all $k\in\mathcal{K}_{r}$ and $\tau\in\mathcal{T}$:
\begin{equation}
\begin{aligned}
&P_{\tau}^{\mathrm{WT}}
-
P_{\tau}^{\mathrm{cur,WT}}
+
P_{\tau}^{\mathrm{PV}}
-
P_{\tau}^{\mathrm{cur,PV}}
\\
&\quad
+
P_{\tau}^{\mathrm{dis}}
-
P_{\tau}^{\mathrm{ch}}
+
P_{\tau}^{\mathrm{buy}}
-
P_{\tau}^{\mathrm{sell}}
=
\tilde{P}_{\tau}^{\mathrm{L}}
-
P_{\tau}^{\mathrm{dlc}}
\end{aligned}
\label{eq:power_balance}
\end{equation}
\begin{equation}
0 \le P_{\tau}^{\mathrm{ch}} \le P_{\mathrm{ESS}}^{\max},
\quad
0 \le P_{\tau}^{\mathrm{dis}} \le P_{\mathrm{ESS}}^{\max}
\label{eq:ess_ch_limit}
\end{equation}
\begin{equation}
E_{\tau+1}
=
E_{\tau}
+
\eta^{\mathrm{ch}} P_{\tau}^{\mathrm{ch}} \Delta t
-
\frac{1}{\eta^{\mathrm{dis}}} P_{\tau}^{\mathrm{dis}} \Delta t
\label{eq:ess_energy_dynamics}
\end{equation}
\begin{equation}
E_{t}=E_{t}^{0}
\label{eq:actual_initial_energy}
\end{equation}
\begin{equation}
0 \le E_{\tau} \le E_{\mathrm{ESS}}^{\max}
\label{eq:ess_energy_limit}
\end{equation}
\begin{equation}
0 \le P_{\tau}^{\mathrm{dlc}} \le \bar{\kappa}\tilde{P}_{\tau}^{\mathrm{L}}
\label{eq:dlc_limit}
\end{equation}
\begin{equation}
0 \le P_{\tau}^{\mathrm{cur,WT}} \le P_{\tau}^{\mathrm{WT}},
\quad
0 \le P_{\tau}^{\mathrm{cur,PV}} \le P_{\tau}^{\mathrm{PV}}
\label{eq:curtail_limit}
\end{equation}
\begin{equation}
0 \le P_{\tau}^{\mathrm{buy}} \le P_{\mathrm{PCC}}^{\max},
\quad
0 \le P_{\tau}^{\mathrm{sell}} \le P_{\mathrm{PCC}}^{\max}
\label{eq:grid_buy_limit}
\end{equation}
\begin{equation}
0 \le \Delta P_{\tau}^{k,+} \le R_{k}^{+},
\quad
0 \le \Delta P_{\tau}^{k,-} \le R_{k}^{-}
\label{eq:reserve_limit}
\end{equation}
where $P_{\tau}^{\mathrm{WT}}$ and $P_{\tau}^{\mathrm{PV}}$ are the known wind and photovoltaic inputs. The curtailment variables in \eqref{eq:power_balance} are introduced to preserve feasibility when the realized load is low while renewable output is high. Equations \eqref{eq:ess_ch_limit}--\eqref{eq:ess_energy_limit} are imposed on the recourse-adjusted actual ESS powers, and the same physical limits are also required for the scheduled variables in the nominal first-stage model. Here, $\bar{\kappa}$ denotes the maximum DLC activation ratio, and $R_{k}^{+}$ and $R_{k}^{-}$ denote the maximum real-time upward and downward adjustment capacities. Both the nominal schedule and the recourse-adjusted operation are initialized from the current ESS state $E_t^{0}$.

To ensure network feasibility, the branch-flow and voltage variables satisfy
\begin{equation}
\left(P_{ij,\tau},Q_{ij,\tau},V_{i,\tau}\right)\in\mathcal{F}_{\mathrm{DistFlow}}
\label{eq:distflow_compact}
\end{equation}
\begin{equation}
1-V^{\max}\le V_{i,\tau}\le 1+V^{\max}
\label{eq:voltage_limit}
\end{equation}
for all buses $i$, branches $(i,j)$, and periods $\tau\in\mathcal{T}$, where $\mathcal{F}_{\mathrm{DistFlow}}$ denotes the feasible set defined by the DistFlow equations for radial distribution networks \cite{baran1989optimal}.

The feasible sets $\mathcal{X}$ and $\mathcal{Y}(x,u)$ in \eqref{eq:tsro_compact} are therefore determined by \eqref{eq:nominal_power_balance}--\eqref{eq:voltage_limit}. The first stage determines a physically feasible base operating schedule under the forecasted load trajectory, while the second stage modifies that schedule through bounded recourse actions after the load realization is observed. In this way, the TSRO model explicitly captures the coupling between planned operation and corrective recourse, while preserving both physical feasibility and engineering realizability. Note that the nominal first-stage schedule is identically constrained by the network feasibility conditions in \eqref{eq:distflow_compact} and \eqref{eq:voltage_limit} evaluated under the forecasted base-case load. The mutual exclusiveness constraints of charging/discharging and buying/selling can be included in the original TSRO with binary variables, and their convex relaxation will be introduced later in the surrogate TSRO model of Section III.

\subsection{Overall Operational Framework}

\begin{figure*}[ht]
\centering
\includegraphics[width=\textwidth]{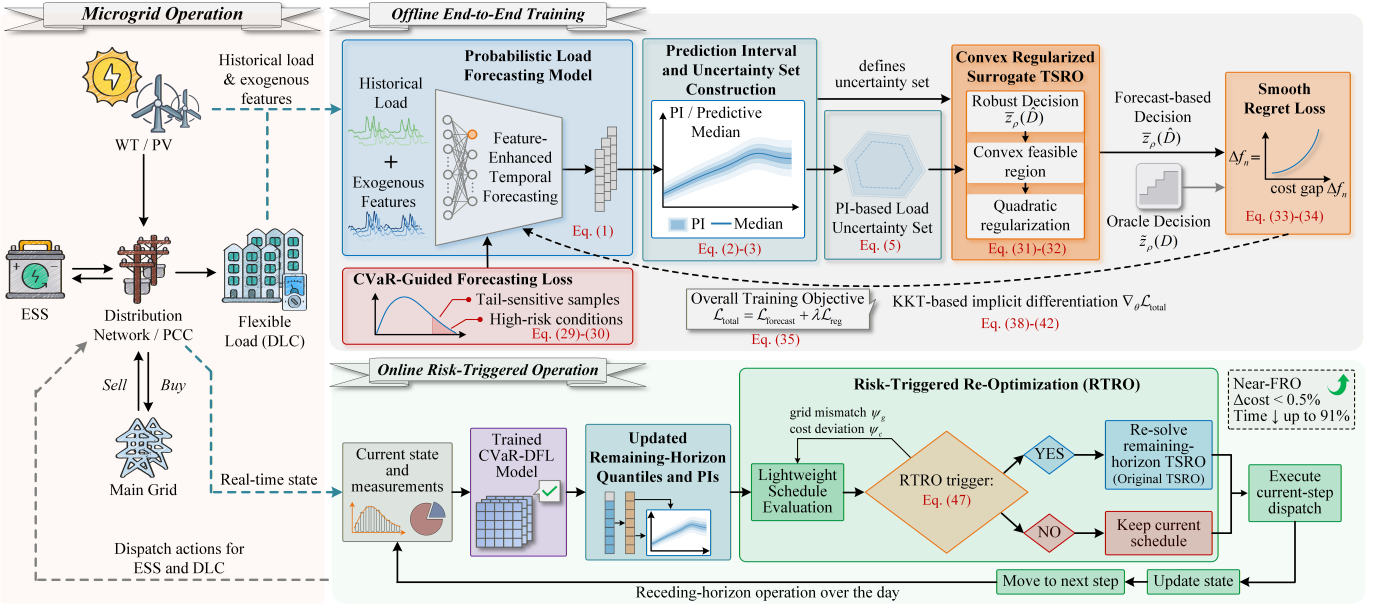}
\caption{Overall framework of the proposed method.}
\label{fig:overallframework}
\vspace{-1.5mm}
\end{figure*}

The overall framework of the proposed method is shown in Fig.~\ref{fig:overallframework}. The upper part of the framework describes the offline forecast-decision coupling between probabilistic load forecasting and robust optimization. Historical load demand and exogenous features are fed into the forecasting model to produce multi-quantile outputs. These quantiles determine the load PI bounds and thus parameterize the uncertainty set of the downstream TSRO model. During training, the forecasting model is optimized not only by forecasting losses, but also by operational feedback from the downstream surrogate TSRO model.

The lower part of the framework corresponds to online microgrid operation. At each operating step, the trained forecasting model updates the remaining-horizon load uncertainty information, and the original TSRO model generates the dispatch schedule considering ESS coordination and DLC recourse. Instead of re-solving the remaining-horizon optimization at every step, the proposed risk-triggered re-optimization mechanism monitors whether the current schedule becomes significantly mismatched with the updated system condition. Re-optimization is activated only when the operational risk indicator exceeds the prescribed threshold or when the elapsed interval since the last update becomes too large. In this way, the proposed framework unifies probabilistic load forecasting, robust dispatch, and computationally efficient online adaptation.

\section{Methodology}
\label{sec:methodology}

\subsection{CVaR-Guided Probabilistic Load Forecasting}

Since the probabilistic load forecasting outputs directly determine the PI-based uncertainty set of the subsequent TSRO model, the forecasting model should not only fit the average prediction error, but also maintain satisfactory performance on those difficult samples that are more influential to robust operation. To this end, a CVaR-guided probabilistic load forecasting method is developed in this paper.

In the case study, the proposed CVaR-DFL model and the standard DFL benchmark share the same probabilistic forecasting backbone, which is adapted from our previous feature-enhanced deep learning architecture for probabilistic forecasting \cite{CAO2024123751}. This backbone maps historical load observations and exogenous features to multi-quantile outputs over the remaining horizon through a feature-enhanced temporal forecasting structure. In the present paper, the original architecture is tailored to short-term microgrid load forecasting and further enhanced by the proposed CVaR-guided objective, so that the model places more emphasis on high-loss and risk-sensitive samples. This design enables the forecasting module to provide uncertainty information that is better aligned with downstream decision-focused training and two-stage robust microgrid operation.

Let $\mathcal{Q}=\{\alpha_{1},\ldots,\alpha_{M}\}$ denote the set of quantile levels to be predicted. For a sample $n$ in a mini-batch $\mathcal{B}$, let $P_{\tau,n}^{\mathrm{L}}$ denote the realized load at future period $\tau\in\mathcal{T}$, and let $\hat{q}_{\tau,n}^{(\alpha)}$ denote the corresponding predicted $\alpha$-quantile, where $\alpha\in\mathcal{Q}$. The quantile prediction error is measured by the pinball loss
\begin{equation}
\ell_{\alpha}(y,\hat{q})
=
\alpha(y-\hat{q})_{+}
+
(1-\alpha)(\hat{q}-y)_{+}
\label{eq:pinball}
\end{equation}
where $(a)_{+}=\max(a,0)$. Based on \eqref{eq:pinball}, the sample-wise aggregated quantile loss is defined as
\begin{equation}
\zeta_{n}
=
\frac{1}{|\mathcal{T}|\,|\mathcal{Q}|}
\sum_{\tau\in\mathcal{T}}
\sum_{\alpha\in\mathcal{Q}}
\ell_{\alpha}
\left(
P_{\tau,n}^{\mathrm{L}},
\hat{q}_{\tau,n}^{(\alpha)}
\right)
\label{eq:sample_quantile_loss}
\end{equation}

Minimizing only the batch-average of \eqref{eq:sample_quantile_loss} mainly improves the average forecasting accuracy, but may insufficiently emphasize those high-loss samples associated with sharp load fluctuations or difficult operating conditions. To enhance the forecasting robustness on such samples, the CVaR of the sample losses is introduced \cite{rockafellar2000optimization}. For a prescribed confidence level $\alpha_{\mathrm{c}}\in(0,1)$, which is distinct from the quantile levels in $\mathcal{Q}$, the CVaR-related objective is written as
\begin{equation}
\mathcal{L}_{\mathrm{CVaR}}(\xi)
=
\xi
+
\frac{1}{(1-\alpha_{\mathrm{c}})|\mathcal{B}|}
\sum_{n\in\mathcal{B}}
(\zeta_{n}-\xi)_{+}
\label{eq:cvar_loss}
\end{equation}
where $\xi$ is an auxiliary scalar variable. During training, $\xi$ is treated as a learnable variable and jointly updated with the forecasting parameters. Equation \eqref{eq:cvar_loss} evaluates the upper tail of the batch loss distribution, and therefore explicitly strengthens the learning emphasis on difficult samples.

Accordingly, the forecasting loss is constructed as
\begin{equation}
\mathcal{L}_{\mathrm{forecast}}
=
(1-\beta)
\frac{1}{|\mathcal{B}|}
\sum_{n\in\mathcal{B}}
\zeta_{n}
+
\beta
\mathcal{L}_{\mathrm{CVaR}}(\xi)
\label{eq:forecast_loss}
\end{equation}
where $\beta\in[0,1]$ is the risk weight. The first term in \eqref{eq:forecast_loss} preserves the overall probabilistic forecasting accuracy, while the second term improves the tail performance of the forecasting model. Therefore, the proposed loss can guide the model to produce load PIs that are accurate on average and more reliable under high-risk operating conditions.

\subsection{Decision-Focused Training via the Surrogate TSRO Model and Regret Loss}

The forecasting model should be trained not only for statistical accuracy, but also for downstream decision quality \cite{doi:10.1287/mnsc.2020.3922}. This paper develops a differentiable surrogate TSRO model to enable forecast-decision coupling during training, with related optimization-layer ideas serving as background \cite{9617122}.

Let $\hat{D}$ denote the forecast output used by the downstream surrogate TSRO model, including the PI bounds generated by the forecasting model. Since direct differentiation of the original TSRO model is not practical due to its binary and non-smooth properties, a convex surrogate TSRO model is constructed by relaxing the binary variables and adding a quadratic regularization term. Let $z$ denote the stacked surrogate decision vector containing the relaxed first-stage and second-stage variables. The forecast-induced surrogate decision is defined as
\begin{equation}
\bar{z}_{\rho}(\hat{D})
=
\arg\min_{z\in\bar{\mathcal{Z}}(\hat{D})}
f_{\mathrm{R}}(z;\hat{D})
+
\frac{\rho}{2}\|z\|_{2}^{2}
\label{eq:surrogate_tsro}
\end{equation}

\begin{figure}[ht]
\centering
\includegraphics[width=\columnwidth]{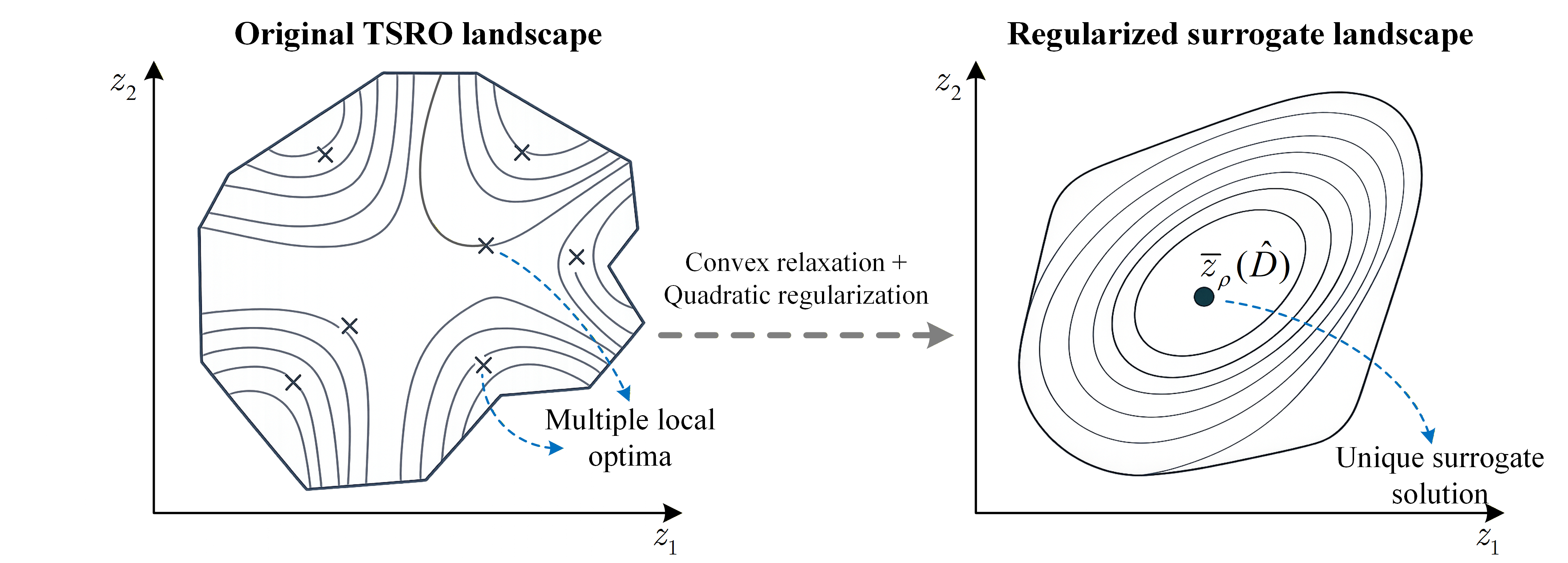}
\caption{Surrogate TSRO illustration.}
\label{fig:surrogate}
\vspace{-1.5mm}
\end{figure}

Fig.~\ref{fig:surrogate} illustrates the surrogate TSRO mechanism. Through convex relaxation and quadratic regularization, the original TSRO landscape is transformed into a surrogate landscape with a unique solution.

In \eqref{eq:surrogate_tsro}, $\bar{\mathcal{Z}}(\hat{D})$ denotes the convex feasible region induced by the forecast output $\hat{D}$, $f_{\mathrm{R}}(z;\hat{D})$ is the surrogate robust objective, and $\rho>0$ is a small quadratic regularization coefficient. The quadratic term ensures strong convexity and a unique surrogate solution. Accordingly, \eqref{eq:surrogate_tsro} is used as the differentiable surrogate TSRO model during training, whereas the final dispatch in online operation and performance evaluation is still obtained by solving the original TSRO model.

For a realized load trajectory $D$, the corresponding oracle decision is defined through a deterministic surrogate dispatch problem in which the uncertainty set degenerates to the realized trajectory. Specifically,
\begin{equation}
\tilde{z}_{\rho}(D)
=
\arg\min_{z\in\tilde{\mathcal{Z}}(D)}
f_{\mathrm{D}}(z;D)
+
\frac{\rho}{2}\|z\|_{2}^{2}
\label{eq:oracle_surrogate}
\end{equation}
where $\tilde{\mathcal{Z}}(D)$ and $f_{\mathrm{D}}(z;D)$ denote the feasible region and objective of the deterministic surrogate dispatch under the realized load trajectory. Therefore, \eqref{eq:surrogate_tsro} generates the forecast-induced robust decision, whereas \eqref{eq:oracle_surrogate} provides the oracle benchmark under perfect information.

Based on the two surrogate TSRO models above, the forecast-induced decision quality is evaluated through the realized cost gap between the forecast-based decision and the oracle decision. For a sample $n$, let $\hat{D}^{(n)}$ and $D^{(n)}$ denote the forecasted and realized load descriptions, respectively. The surrogate decision regret is defined as
\begin{equation}
\Delta f_{n}
=
f_{\mathrm{D}}\!\left(\bar{z}_{\rho}(\hat{D}^{(n)});D^{(n)}\right)
-
f_{\mathrm{D}}\!\left(\tilde{z}_{\rho}(D^{(n)});D^{(n)}\right)
\label{eq:regret_gap}
\end{equation}
where a smaller regret value indicates that the decision induced by the forecasting model is closer to the oracle decision.

To obtain a smooth training objective, the regret loss is constructed in the softplus form as
\begin{equation}
\mathcal{L}_{\mathrm{reg}}
=
\frac{1}{|\mathcal{B}|}
\sum_{n\in\mathcal{B}}
\log
\left(
1+\exp(\Delta f_{n})
\right)
\label{eq:regret_loss}
\end{equation}
Accordingly, the overall training objective is written as
\begin{equation}
\mathcal{L}_{\mathrm{total}}
=
\mathcal{L}_{\mathrm{forecast}}
+
\lambda \mathcal{L}_{\mathrm{reg}}
\label{eq:total_loss}
\end{equation}
where $\mathcal{L}_{\mathrm{forecast}}$ is the forecasting loss defined in \eqref{eq:forecast_loss}, and $\lambda>0$ is the weight of the decision-focused term. In this way, the forecasting model is trained by jointly considering probabilistic forecasting performance and downstream decision quality.

\subsection{Implicit Differentiation of the Surrogate TSRO Model}

The remaining task is to back-propagate the gradient of the decision-focused loss through the forecast-induced surrogate TSRO model in \eqref{eq:surrogate_tsro}. Since the surrogate problem is convex and strongly regularized, its optimal solution can be differentiated through the KKT conditions \cite{agrawal2019differentiable}.

Let the equality and inequality constraints of the surrogate problem be written in the generic form
\begin{equation}
a(z;\hat{D})=0
\label{eq:eq_constraint}
\end{equation}
\begin{equation}
g(z;\hat{D})\le 0
\label{eq:ineq_constraint}
\end{equation}
and let $\nu$ and $\mu$ denote the corresponding Lagrange multipliers. Define $w=(z,\mu,\nu)$ as the stacked primal-dual variable. Then, the KKT residual of the forecast-induced surrogate problem can be expressed as
\begin{equation}
F(w;\hat{D})
=
\begin{bmatrix}
\nabla_{z}f_{\mathrm{R}}(z;\hat{D})+\rho z+J_{a}^{\top}\nu+J_{g}^{\top}\mu\\
a(z;\hat{D})\\
\mu\circ g(z;\hat{D})
\end{bmatrix}
=
0
\label{eq:kkt_residual}
\end{equation}
where $J_{a}=\partial a(z;\hat{D})/\partial z$, $J_{g}=\partial g(z;\hat{D})/\partial z$, and $\circ$ denotes the Hadamard product. Together with the primal feasibility condition in \eqref{eq:ineq_constraint}, the KKT system also requires the dual feasibility condition $\mu \ge 0$.

Under standard regularity conditions, including constraint qualification, strict complementarity, and second-order sufficiency, the complete KKT system associated with the surrogate TSRO model is locally nonsingular in a neighborhood of the optimum. Differentiating \eqref{eq:kkt_residual} with respect to $\hat{D}$ gives
\begin{equation}
\frac{\partial F}{\partial w}
\frac{\partial w}{\partial \hat{D}}
+
\frac{\partial F}{\partial \hat{D}}
=
0
\label{eq:implicit_system}
\end{equation}

To avoid explicitly computing $\partial w/\partial \hat{D}$, an adjoint variable $v$ is introduced such that
\begin{align}
\left(\frac{\partial F}{\partial w}\right)^{\top}v
&=
\left(\frac{\partial \mathcal{L}_{\mathrm{reg}}}{\partial w}\right)^{\top}
\label{eq:adjoint_system}\\
\nabla_{\hat{D}}\mathcal{L}_{\mathrm{reg}}
&=
-
\left(\frac{\partial F}{\partial \hat{D}}\right)^{\top}v
\label{eq:grad_forecast_output}\\
\nabla_{\theta}\mathcal{L}_{\mathrm{total}}
&=
\nabla_{\theta}\mathcal{L}_{\mathrm{forecast}}
+
\lambda
\left(\frac{\partial \hat{D}_{\theta}}{\partial \theta}\right)^{\top}
\nabla_{\hat{D}}\mathcal{L}_{\mathrm{reg}}
\label{eq:grad_theta}
\end{align}

Thus, \eqref{eq:kkt_residual}--\eqref{eq:grad_theta} establish a differentiable forecast-decision pathway for training. During back-propagation, the oracle term in \eqref{eq:regret_gap} is treated as constant, and gradients are propagated only through the forecast-induced surrogate decision. Therefore, the forecasting model is updated according to both statistical forecasting errors and the downstream decision quality induced by its PI outputs.

\subsection{Risk-Triggered Re-Optimization}

A full re-optimization (FRO) policy re-solves the remaining-horizon TSRO at every online step. Although such a policy provides high adaptability, it also incurs substantial online computational cost in repeated online optimization. To reduce unnecessary re-optimizations while retaining most of the economic benefit, a risk-triggered selective receding-horizon re-optimization mechanism, referred to as RTRO, is developed in this paper.

To avoid solving a new TSRO merely for trigger evaluation, the trigger indicators are computed from a lightweight schedule-evaluation step rather than from a full re-optimization. Specifically, at online step $t$, the current remaining-horizon schedule is kept fixed, and the updated remaining-horizon load forecast is injected into the power-balance and cost expressions to estimate the resulting grid exchange and operating cost. Therefore, the quantities used in the trigger rule serve only as screening indicators, whereas the exact remaining-horizon TSRO is solved only after the trigger condition is satisfied.

Let $P_{t}^{\mathrm{grid,sch}}$ denote the net grid-exchange power implied by the current schedule, and let $P_{t}^{\mathrm{grid,upd}}$ denote the corresponding value estimated by the above schedule-evaluation step using the updated system state and the updated remaining-horizon load forecast. Here, the net grid exchange is defined as $P_{t}^{\mathrm{grid}}=P_{t}^{\mathrm{buy}}-P_{t}^{\mathrm{sell}}$. A grid-side mismatch indicator is defined as
\begin{equation}
\psi_{t}^{g}
=
\frac{
\left|
P_{t}^{\mathrm{grid,upd}}-P_{t}^{\mathrm{grid,sch}}
\right|
}{
P_{\mathrm{PCC}}^{\max}
}
\label{eq:grid_indicator}
\end{equation}

Let $\hat{J}_{t}^{\mathrm{sch}}$ denote the estimated remaining-horizon operating cost under the current schedule, and let $\hat{J}_{t}^{\mathrm{upd}}$ denote the corresponding cost obtained by the same schedule-evaluation step under the updated system state and the updated remaining-horizon load forecast. A cost-side deviation indicator is defined as
\begin{equation}
\psi_{t}^{c}
=
\frac{
\left|
\hat{J}_{t}^{\mathrm{upd}}-\hat{J}_{t}^{\mathrm{sch}}
\right|
}{
\left|
\hat{J}_{t}^{\mathrm{sch}}
\right|+\varepsilon
}
\label{eq:cost_indicator}
\end{equation}
where $\varepsilon$ is a small positive constant to avoid division by zero. The overall normalized risk indicator is then written as
\begin{equation}
\psi_{t}
=
\max
\left\{
\frac{\psi_{t}^{g}}{\epsilon_{g}},
\frac{\psi_{t}^{c}}{\epsilon_{c}}
\right\}
\label{eq:combined_indicator}
\end{equation}
where $\epsilon_{g}$ and $\epsilon_{c}$ are the trigger thresholds for the grid-side and cost-side deviations, respectively.

Let $t_{\mathrm{last}}$ denote the most recent time step at which re-optimization was executed. The elapsed interval since the last re-optimization is defined as
\begin{equation}
\Delta\tau_{t}=t-t_{\mathrm{last}}
\label{eq:elapsed_interval}
\end{equation}

Based on \eqref{eq:combined_indicator} and \eqref{eq:elapsed_interval}, the RTRO trigger rule is designed as
\begin{equation}
\chi_{t}
=
\begin{cases}
1, & \left(\psi_{t}>1\ \wedge\ \Delta\tau_{t}\ge \Delta\tau_{\min}\right)\ \vee\ \left(\Delta\tau_{t}\ge \Delta\tau_{\max}\right) \\
0, & \text{otherwise}
\end{cases}
\label{eq:trigger_rule}
\end{equation}
where $\chi_{t}=1$ indicates that the remaining-horizon TSRO is re-solved at step $t$.

The physical meaning of \eqref{eq:trigger_rule} is straightforward. The lower window $\Delta\tau_{\min}$ suppresses excessively frequent re-optimization caused by temporary fluctuations, whereas the upper window $\Delta\tau_{\max}$ prevents the schedule from becoming outdated for too long. The threshold $\epsilon_{g}$ filters out negligible grid-side deviations, and $\epsilon_{c}$ filters out small cost variations that do not justify a full re-optimization. Once the trigger condition is satisfied, the exact remaining-horizon TSRO is re-solved using the updated system state and the updated load forecast, and the unexecuted part of the previous schedule is replaced by the new solution. Otherwise, the current schedule continues to be executed. In this way, RTRO uses low-cost screening indicators to approximate the need for re-optimization, and thus preserves near-FRO operating performance while substantially reducing online computational cost.

\subsection{Overall Algorithm and Workflow}

Algorithm~\ref{alg:main} summarizes the offline training and online deployment of the proposed method. In the offline stage, the forecasting model is trained by jointly minimizing the forecasting loss and the decision-focused regret through the surrogate TSRO model. In the online stage, the trained model updates the remaining-horizon load quantiles and the PI-based uncertainty set, while RTRO selectively re-solves the original TSRO only when the trigger condition is satisfied.

\begin{algorithm}[ht]
\caption{Offline training and online operation of the proposed method}
\label{alg:main}
\footnotesize
\begin{algorithmic}[1]
\STATE \textbf{Input:} Training set $\{(\mathbf{X}^{(n)},D^{(n)})\}_{n=1}^{N}$, initialized forecasting parameter $\theta$, auxiliary scalar $\xi$, CVaR level $\alpha_{\mathrm{c}}$, risk weight $\beta$, regret weight $\lambda$, regularization coefficient $\rho$
\FOR{epoch $=1$ to $E$}
    \STATE Sample a mini-batch $\mathcal{B}$, generate $\hat{D}_{\theta}^{(n)}$, and for each $n\in\mathcal{B}$ compute $\zeta_n$, $\bar{z}_{\rho}(\hat{D}_{\theta}^{(n)})$, $\tilde{z}_{\rho}(D^{(n)})$, and $\Delta f_n$
    \STATE Evaluate $\mathcal{L}_{\mathrm{forecast}}$, $\mathcal{L}_{\mathrm{reg}}$, and $\mathcal{L}_{\mathrm{total}}$ by \eqref{eq:forecast_loss}--\eqref{eq:total_loss}, then update $\theta$ using \eqref{eq:adjoint_system}--\eqref{eq:grad_theta} and update $\xi$ by minimizing \eqref{eq:cvar_loss}
\ENDFOR
\STATE \textbf{Online operation:} Initialize the system state, generate the initial remaining-horizon forecast, solve the original TSRO over the full horizon, and set $t_{\mathrm{last}}=1$
\STATE Execute the decision for the first interval
\FOR{$t=2$ to $T$}
    \STATE Update the current state and remaining-horizon forecast, and compute $\psi_t^{g}$, $\psi_t^{c}$, $\psi_t$, and $\Delta\tau_t$
    \IF{$\chi_t=1$ according to \eqref{eq:trigger_rule}}
        \STATE Re-solve the remaining-horizon TSRO, update the unexecuted schedule, and set $t_{\mathrm{last}}=t$
    \ENDIF
    \STATE Execute the decision for interval $t$
\ENDFOR
\end{algorithmic}
\end{algorithm}

\begin{figure}[ht]
\centering
\includegraphics[width=8.7cm]{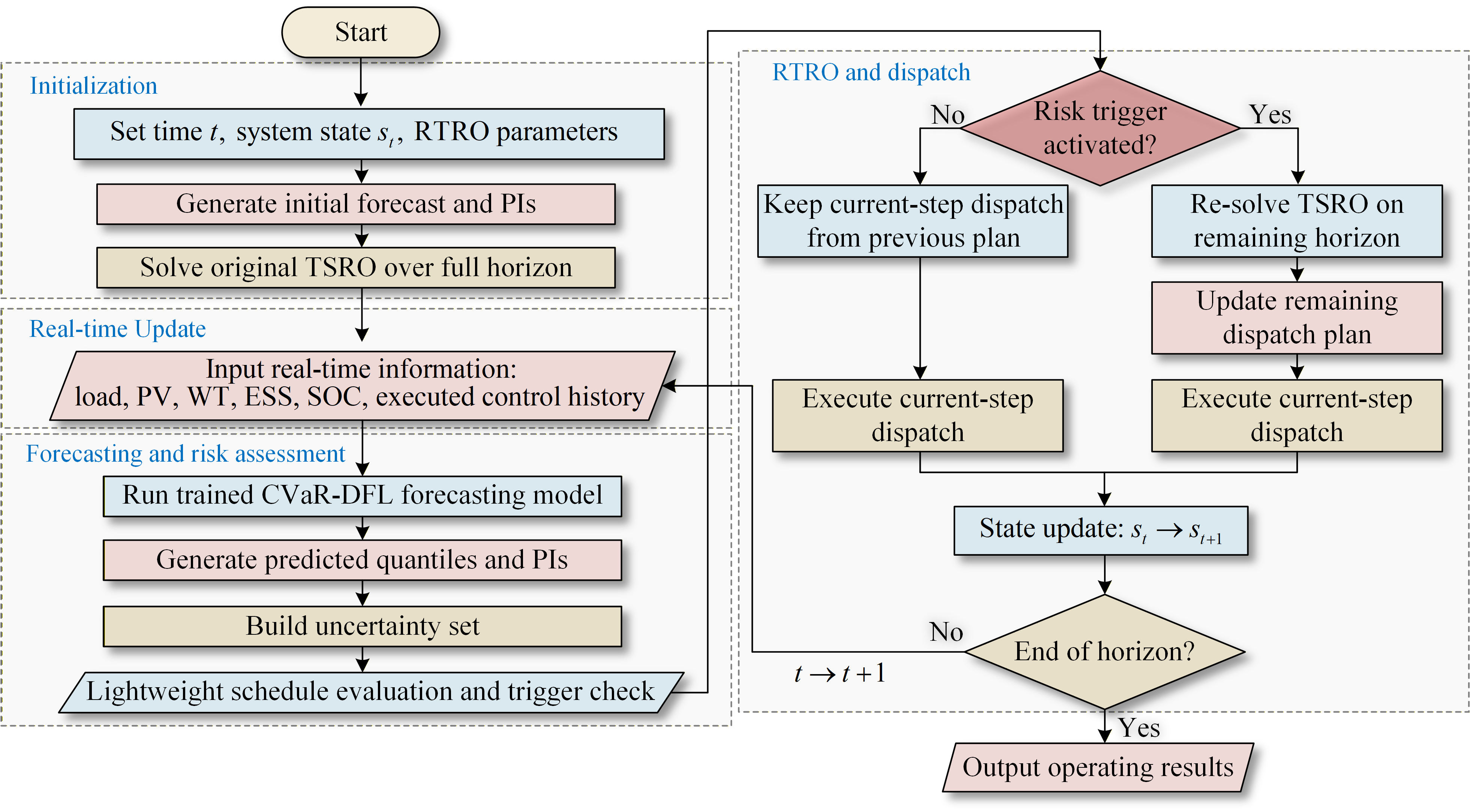}
\caption{Online operational workflow of the proposed method.}
\label{fig:workflow}
\vspace{-1.5mm}
\end{figure}

Fig.~\ref{fig:workflow} focuses on the online operational workflow after offline training. The procedure starts from initialization, where the current time, system state, and RTRO parameters are set, the initial forecast and PIs are generated, and the original TSRO is solved over the full horizon. During operation, real-time information is collected to update the system state, the trained CVaR-DFL model produces the remaining-horizon quantiles and PIs, and a lightweight schedule-evaluation step computes the trigger indicators. If the risk trigger is activated, the remaining-horizon TSRO is re-solved and the unexecuted dispatch plan is updated; otherwise, the current-step action from the previous plan is retained. The current-step dispatch is then executed, the state is updated, and the same procedure repeats until the end of the horizon.

\section{Case Study}
\label{sec:case_study}

\subsection{Experimental Setup}
\label{subsec:exp_setup}
\begin{figure}[ht]
\centering
\includegraphics[width=8.4cm]{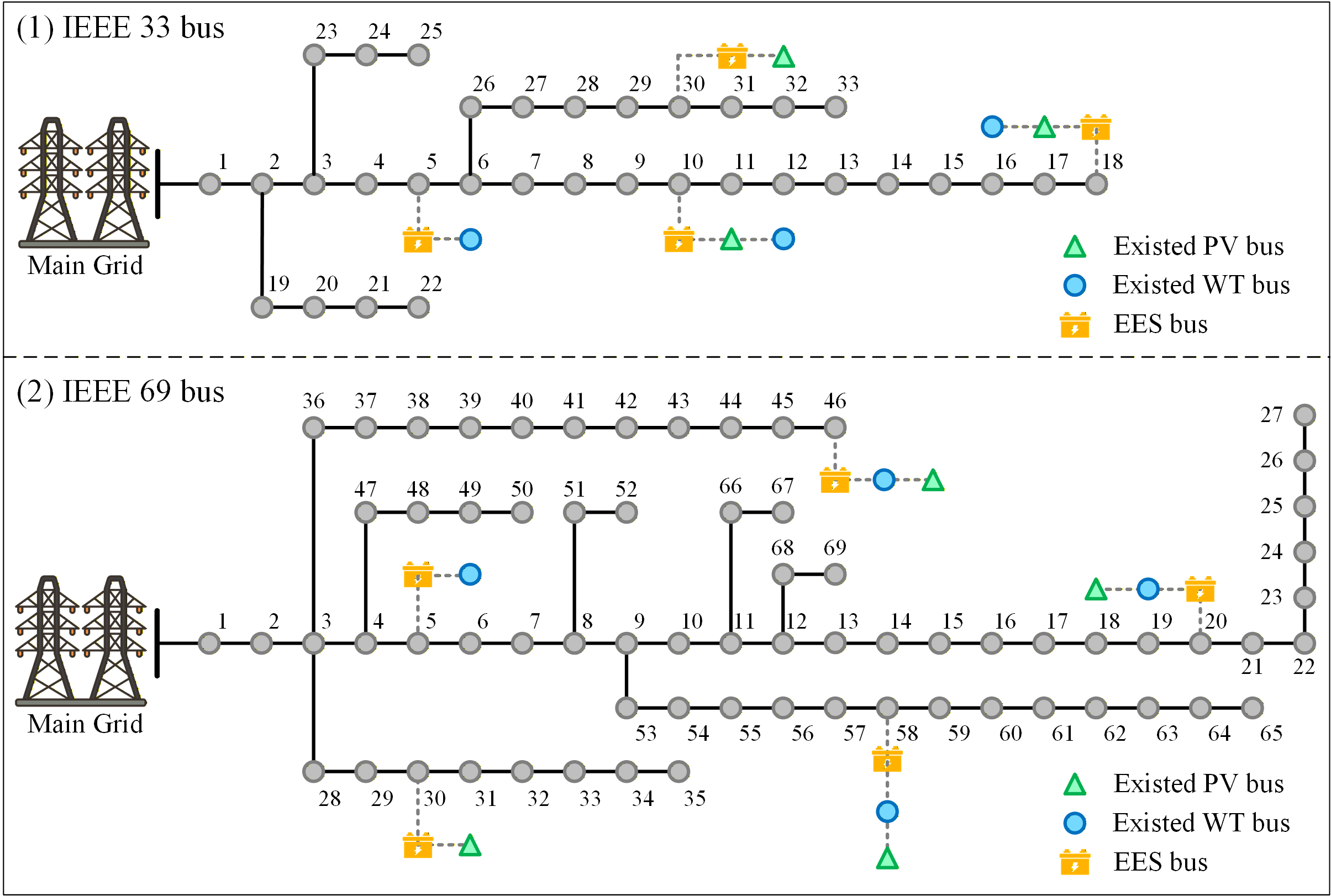}
\caption{IEEE 33-bus and 69-bus microgrid topologies.}
\label{fig:iv01_testbed}
\vspace{-1.5mm}
\end{figure}

The proposed method is tested on modified IEEE 33-bus and 69-bus grid-connected microgrids with wind generation, photovoltaic generation, an ESS, DLC, and power exchange with the main grid as shown in Fig.~\ref{fig:iv01_testbed}. The dispatch interval is 15 min and the operation horizon is 24 h, resulting in 96 decision steps per day. Accordingly, both the forecasting horizon and the TSRO dispatch horizon are implemented on a 15-min time grid in the case study. In the case study, wind and photovoltaic outputs are treated as known inputs over the dispatch horizon. The key system and method parameters are summarized in Table~\ref{tab:key_params}.

\begin{table}[ht]
\caption{Key system and method parameters.}
\label{tab:key_params}
\centering
\renewcommand{\arraystretch}{1.3}
\scriptsize
\begin{tabular}{l c}
\hline
\textbf{Parameter} & \textbf{Value} \\
\hline
$P^{\max}_{\mathrm{PCC}}$ (grid exchange power limit) & 5000 kW \\
$P^{\max}_{\mathrm{ESS}}$ (charge/discharge power limit) & 1100 kW \\
$E^{\max}_{\mathrm{ESS}}$ (ESS energy capacity) & 6000 kWh \\
$P^{\max}_{\mathrm{PV}}$ & 3200 kW \\
$P^{\max}_{\mathrm{WT}}$ & 1000 kW \\
\hline
$\alpha_{\mathrm{c}}$ (CVaR level) & 0.95 \\
$\beta$ (risk weight) & 0.50 \\
$\lambda$ (loss weight) & 1.00 \\
$\rho$ (surrogate regularization) & $10^{-3}$ \\
$\Delta\tau_{\min}, \Delta\tau_{\max}$ (RTRO window) & $1,\ 8$ \\
$\epsilon_g, \epsilon_c$ (trigger thresholds) & $10^{-3},\ 0.05$ \\
\hline
\end{tabular}
\vspace{-1.5mm}
\end{table}

For the forecasting task, a real-world historical dataset with 15-min resolution from Jan.~1,~2022 to Dec.~31,~2022 is utilized. For each target day, the model is trained on the previous 14 days and used for next-day 24 h forecasting. The predictive median is taken as the nominal load forecast, and the corresponding 90\% PI is used to construct the uncertainty set of the downstream TSRO model. Two representative operating conditions are selected for detailed illustration, namely a typical day and an extreme day, where the latter is characterized by higher load volatility and larger forecasting difficulty.

The proposed method is compared with long short-term memory (LSTM) \cite{kong2017short}, Transformer \cite{eisenach2020mqtransformer}, and conventional DFL benchmarks. The LSTM and Transformer baselines are accuracy-driven probabilistic forecasting models, whereas the DFL baseline omits the proposed CVaR-guided training. In the online operation study, the FRO baseline is compared with the proposed RTRO strategy for selective remaining-horizon re-optimization. Forecasting performance is evaluated by root mean square error (RMSE), continuous ranked probability score (CRPS), prediction interval coverage probability (PICP), and Pinball loss. Online operation performance is assessed by operating cost, $\mathrm{CVaR}_{90\%}$, average solution time per solve, and total daily solution time.

All forecasting models are implemented in PyTorch, and Gurobi is used as the optimization solver. The experiments are conducted on a workstation with an AMD Ryzen 7 9700X 8-core CPU, 32 GB RAM, and an NVIDIA RTX 4070 Ti Super GPU.

\subsection{Probabilistic Load Forecasting Illustration}
\label{subsec:forecast_illustration}

\begin{figure}[ht]
\centering
\subfloat[Typical day.]{%
    \includegraphics[width=8.4cm]{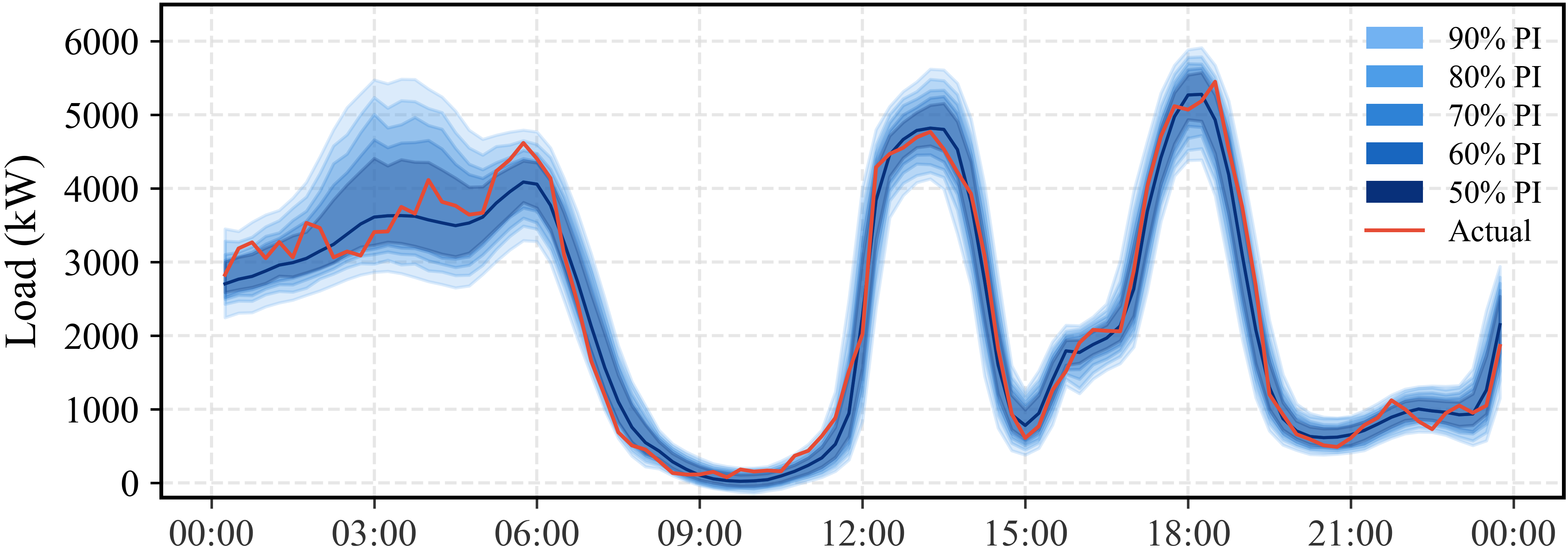}
    \label{fig:typical_forecast}
}\\[0.5mm]
\subfloat[Extreme day.]{%
    \includegraphics[width=8.4cm]{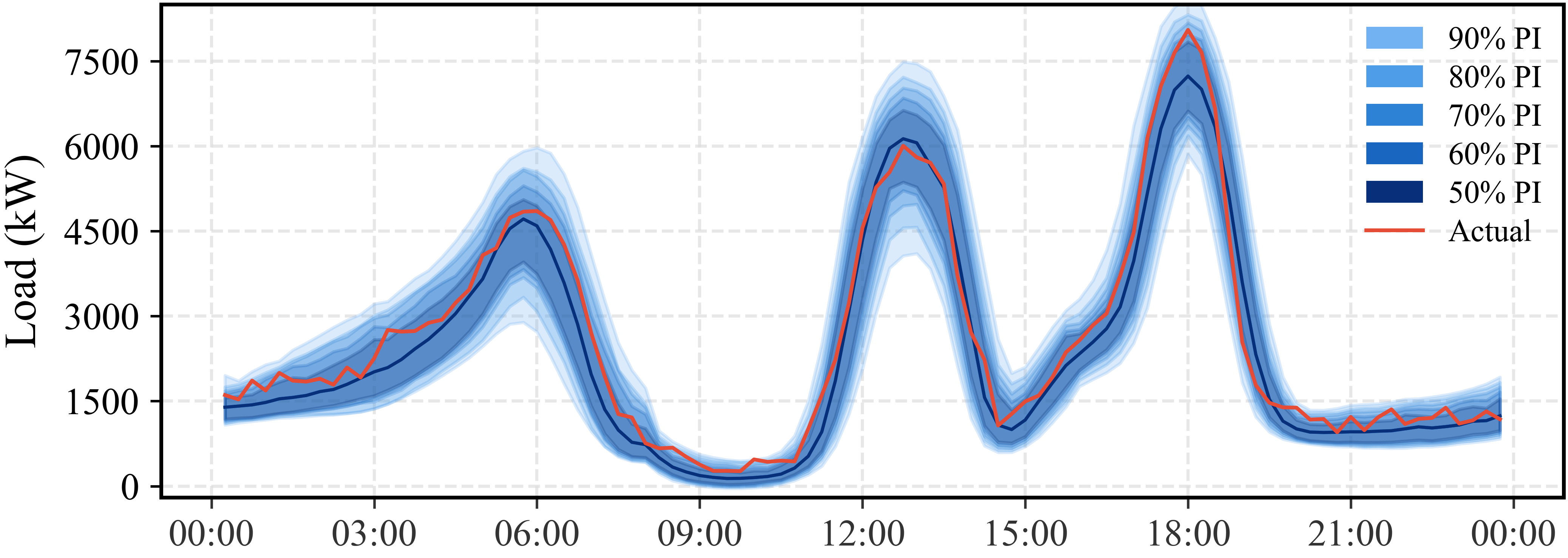}
    \label{fig:extreme_forecast}
}\\[2mm]
\caption{Probabilistic load forecasting results on representative typical and extreme days.}
\label{fig:forecast}
\vspace{-1.5mm}
\end{figure}

Fig.~\ref{fig:forecast} shows the probabilistic load forecasting results on the representative typical and extreme days. On the typical day, the predicted median follows the main load trajectory well and the prediction intervals remain relatively compact over most periods. On the extreme day, sharper ramps and higher peaks lead to visibly wider intervals around the more volatile periods. In both cases, the actual load trajectory is well captured by the predicted intervals, indicating that the proposed model provides effective uncertainty quantification in addition to accurate point prediction.

Table~\ref{tab:forecasting} reports the quantitative forecasting performance. The proposed CVaR-DFL model achieves the best overall results on both representative days, with the lowest RMSE, CRPS, and Pinball loss among all methods. Compared with the conventional DFL model, its improvement is more evident on the extreme day in interval-oriented metrics, where $\mathrm{PICP}_{90\%}$ increases from 0.94 to 0.99 and the Pinball loss decreases from 72.85 to 67.53 kW. This is consistent with the role of the CVaR-guided term, which places more emphasis on tail-risk-sensitive forecasting under stressed operating conditions.

The $\mathrm{PICP}_{90\%}$ values of the proposed model are above the nominal 90\% level, especially on the typical day. Although this may suggest conservative intervals from a purely statistical perspective, such a tendency is more acceptable in the present decision-focused setting, where interval under-coverage can underestimate load uncertainty and reduce downstream decision quality. Meanwhile, the proposed model also achieves lower CRPS, Pinball loss, and RMSE, indicating that the performance gain is not merely caused by wider intervals.

\begin{table*}[ht]
\caption{Probabilistic load forecasting performance on the representative typical and extreme days.}
\label{tab:forecasting}
\centering
\renewcommand{\arraystretch}{1.3}
\scriptsize
\begin{tabular}{lcccc|cccc}
\hline
\multirow{2}{*}{Method} &
\multicolumn{4}{c|}{Typical day} &
\multicolumn{4}{c}{Extreme day} \\
\cline{2-9}
& RMSE (kW) & CRPS (kW) & $\text{PICP}_{90\%}$ & $\text{Pinball}_{90\%}$ (kW)
& RMSE (kW) & CRPS (kW) & $\text{PICP}_{90\%}$ & $\text{Pinball}_{90\%}$ (kW) \\
\hline
LSTM         & 316.84 & 188.76 & 0.91 & 68.42
             & 445.92 & 278.63 & 0.88 & 88.71 \\
Transformer  & 299.57 & 178.34 & 0.93 & 64.91
             & 424.38 & 257.46 & 0.91 & 80.36 \\
DFL          & 281.63 & 166.88 & 0.95 & 60.74
             & 398.74 & 238.91 & 0.94 & 72.85 \\
CVaR-DFL (proposed) & 267.32 & 158.44 & 1.00 & 57.69
             & 378.65 & 226.82 & 0.99 & 67.53 \\
\hline
\end{tabular}
\vspace{-1mm}
\end{table*}

\begin{table*}[ht]
\caption{Online operation performance on the extreme day.}
\label{tab:online}
\centering
\renewcommand{\arraystretch}{1.3}
\scriptsize
\begin{tabular}{lcccc|cccc}
\hline
\multirow{2}{*}{Method} &
\multicolumn{4}{c|}{IEEE 33-bus} &
\multicolumn{4}{c}{IEEE 69-bus} \\
\cline{2-9}
& Cost (\$) & $\mathrm{CVaR}_{90\%}$ & Solve time (s) & Total time (min)
& Cost (\$) & $\mathrm{CVaR}_{90\%}$ & Solve time (s) & Total time (min) \\
\hline
LSTM-FRO               & 4516.8 & 4832.4 & 27.17 & 43.47
                      & 7014.6 & 7628.3 & 42.44 & 67.90 \\
Transformer-FRO        & 4392.7 & 4706.3 & 29.65 & 47.44
                      & 6862.1 & 7421.5 & 47.18 & 75.49 \\
DFL-FRO                & 4297.4 & 4608.6 & 32.13 & 51.41
                      & 6719.3 & 7274.8 & 54.94 & 87.90 \\
CVaR-DFL-FRO           & 4174.9 & 4448.7 & 37.81 & 60.49
                      & 6528.7 & 7031.6 & 60.03 & 96.05 \\
CVaR-DFL-RTRO (proposed) & 4188.6 & 4461.9 & 40.18 & 5.29
                      & 6549.6 & 7052.8 & 69.34 & 12.54 \\
\hline
\end{tabular}
\vspace{-1.5mm}
\end{table*}

\begin{table*}[ht]
\caption{Multi-day statistical online operation performance over different test subsets (mean$\pm$std over daily runs).}
\label{tab:online_stats}
\centering
\renewcommand{\arraystretch}{1.3}
\scriptsize
\begin{tabular}{llccc|ccc}
\hline
\multirow{2}{*}{Subset} & \multirow{2}{*}{Method} &
\multicolumn{3}{c|}{IEEE 33-bus} &
\multicolumn{3}{c}{IEEE 69-bus} \\
\cline{3-8}
& & Cost (\$) & $\mathrm{CVaR}_{90\%}$ & Total time (min/day)
& Cost (\$) & $\mathrm{CVaR}_{90\%}$ & Total time (min/day) \\
\hline

\multirow{5}{*}{Typical-day set}
& LSTM-FRO                  & 3812.6$\pm$146.8 & 4026.9$\pm$165.4 & 41.83$\pm$2.21 & 5928.4$\pm$218.6 & 6374.2$\pm$247.3 & 65.81$\pm$3.45 \\
& Transformer-FRO          & 3698.4$\pm$139.7 & 3907.1$\pm$158.6 & 45.76$\pm$2.49 & 5792.7$\pm$209.8 & 6210.5$\pm$238.1 & 72.96$\pm$3.86 \\
& DFL-FRO                  & 3609.7$\pm$131.5 & 3798.6$\pm$151.2 & 49.82$\pm$2.87 & 5668.3$\pm$201.4 & 6078.9$\pm$229.4 & 84.76$\pm$4.72 \\
& CVaR-DFL-FRO             & 3502.8$\pm$126.4 & 3669.5$\pm$144.8 & 58.63$\pm$3.41 & 5519.6$\pm$193.7 & 5889.8$\pm$219.2 & 92.68$\pm$5.21 \\
& CVaR-DFL-RTRO (proposed) & 3516.9$\pm$129.3 & 3685.7$\pm$147.6 &  4.34$\pm$1.28 & 5541.2$\pm$197.1 & 5912.7$\pm$223.6 & 10.46$\pm$3.08 \\
\hline

\multirow{5}{*}{Extreme-day set}
& LSTM-FRO                  & 4511.3$\pm$182.7 & 4824.1$\pm$201.6 & 43.62$\pm$2.56 & 7008.9$\pm$267.5 & 7619.7$\pm$298.4 & 68.14$\pm$4.12 \\
& Transformer-FRO          & 4389.5$\pm$174.1 & 4701.8$\pm$192.4 & 47.58$\pm$2.83 & 6857.6$\pm$258.2 & 7413.6$\pm$286.7 & 75.86$\pm$4.58 \\
& DFL-FRO                  & 4294.2$\pm$166.3 & 4602.9$\pm$185.7 & 51.76$\pm$3.09 & 6714.8$\pm$249.1 & 7268.5$\pm$274.3 & 88.27$\pm$5.11 \\
& CVaR-DFL-FRO             & 4179.6$\pm$158.8 & 4453.5$\pm$176.2 & 60.71$\pm$3.84 & 6534.1$\pm$241.8 & 7036.9$\pm$261.2 & 96.48$\pm$5.74 \\
& CVaR-DFL-RTRO (proposed) & 4192.8$\pm$163.4 & 4468.2$\pm$180.5 &  5.61$\pm$1.74 & 6557.9$\pm$245.6 & 7064.8$\pm$265.9 & 13.27$\pm$4.11 \\
\hline

\multirow{5}{*}{All test days}
& LSTM-FRO                  & 4019.4$\pm$301.7 & 4268.5$\pm$332.8 & 42.27$\pm$2.43 & 6425.7$\pm$487.2 & 6914.6$\pm$522.5 & 66.92$\pm$3.88 \\
& Transformer-FRO          & 3896.8$\pm$294.1 & 4140.7$\pm$323.6 & 46.20$\pm$2.71 & 6291.4$\pm$472.8 & 6748.3$\pm$507.6 & 74.21$\pm$4.31 \\
& DFL-FRO                  & 3805.5$\pm$286.3 & 4035.9$\pm$311.4 & 50.31$\pm$3.02 & 6168.5$\pm$459.7 & 6612.4$\pm$493.8 & 86.03$\pm$4.89 \\
& CVaR-DFL-FRO             & 3697.2$\pm$279.8 & 3904.4$\pm$298.7 & 59.28$\pm$3.66 & 6015.8$\pm$445.9 & 6427.1$\pm$479.4 & 94.57$\pm$5.36 \\
& CVaR-DFL-RTRO (proposed) & 3711.8$\pm$283.1 & 3919.6$\pm$302.5 &  4.78$\pm$1.63 & 6038.6$\pm$452.4 & 6451.5$\pm$484.8 & 11.72$\pm$3.64 \\
\hline
\end{tabular}%
\end{table*}

\begin{figure}[ht]
\centering

\subfloat[RTRO activation.]{%
    \includegraphics[width=8.2cm]{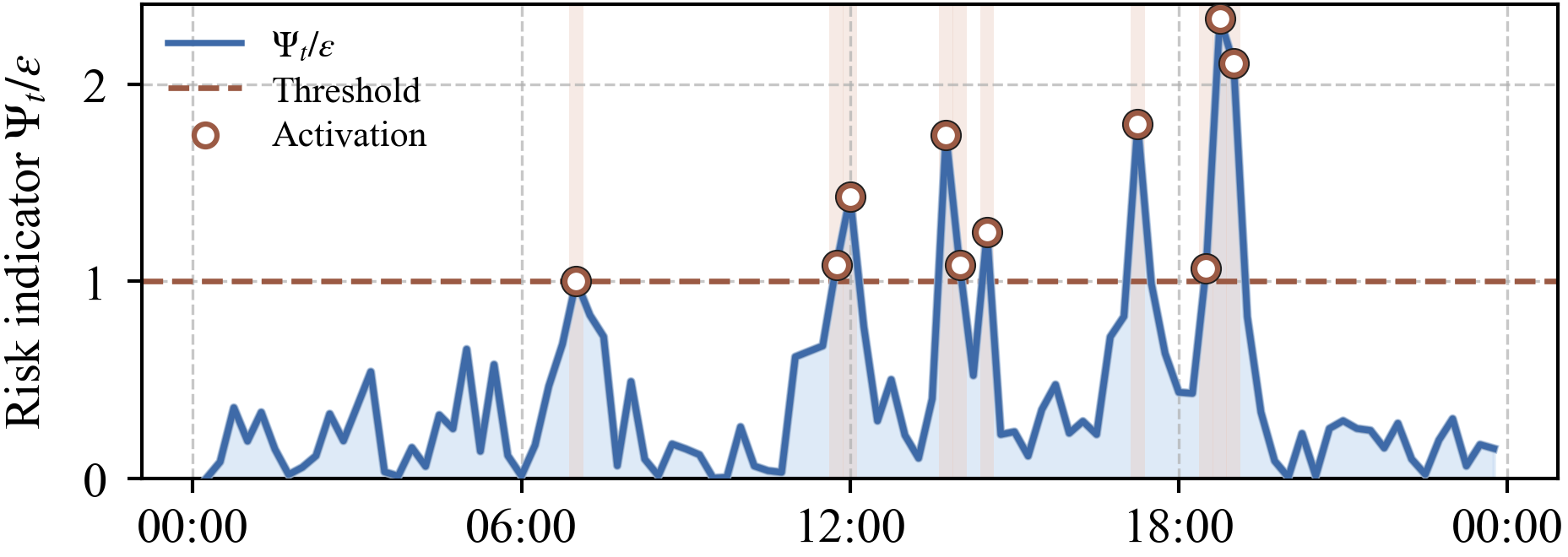}\vspace{3pt}
    \label{fig:RTRO_a}
}\\[2mm]

\subfloat[Per-solve time.]{%
    \includegraphics[width=8.4cm]{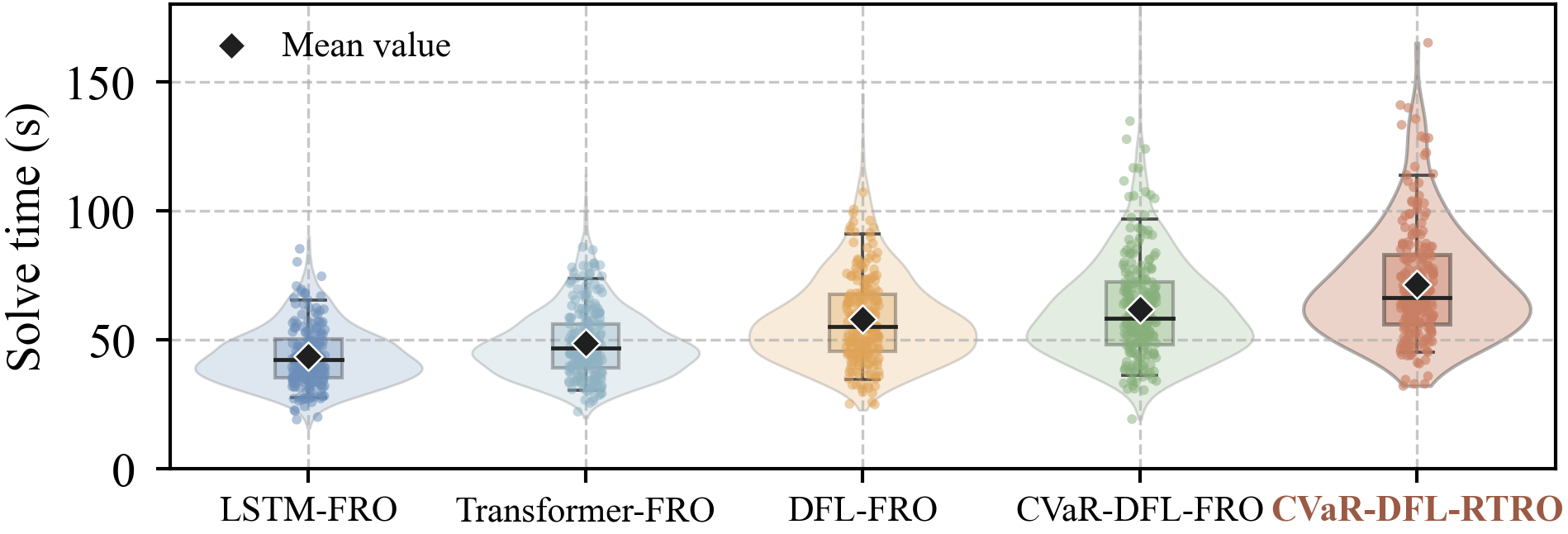}\vspace{3pt}
    \label{fig:RTRO_b}
}\\[2mm]

\subfloat[Daily total time and re-optimization frequency.]{%
    \includegraphics[width=8.4cm]{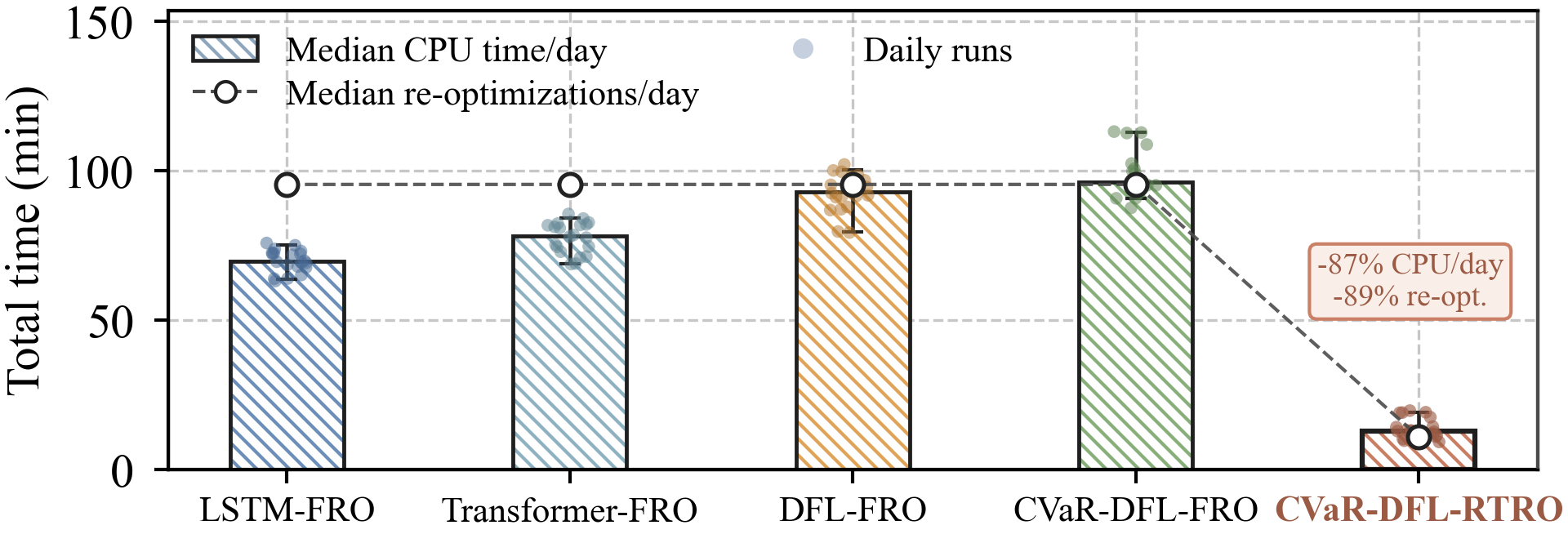}\vspace{3pt}
    \label{fig:RTRO_c}
}\\[2mm]
\caption{RTRO activation and computational performance of different methods.}
\label{fig:RTRO}
\vspace{-1.5mm}
\end{figure}

\subsection{Online Operation Performance and RTRO Effectiveness}
\label{subsec:online_performance}

Table~\ref{tab:online} reports the online operation performance on the representative extreme day. Among the FRO methods, CVaR-DFL-FRO achieves the lowest operating cost and $\mathrm{CVaR}_{90\%}$ on both the IEEE 33-bus and 69-bus systems. Compared with DFL-FRO, it further reduces the operating cost by \$122.5 and \$190.6, respectively, while also lowering $\mathrm{CVaR}_{90\%}$, indicating that the CVaR-guided probabilistic load forecasting model provides more effective uncertainty information for the downstream TSRO model. Compared with CVaR-DFL-FRO, the proposed CVaR-DFL-RTRO causes only marginal increases in cost and risk, but reduces the daily total solution time from 60.49 to 5.29 min on the IEEE 33-bus system and from 96.05 to 12.54 min on the IEEE 69-bus system. This indicates that RTRO preserves near-FRO operational quality while greatly reducing the online computational cost.

Fig.~\ref{fig:RTRO} further explains the source of this gain. RTRO is activated only at limited high-risk periods rather than at every time step. The per-solve time of CVaR-DFL-based methods is comparable to, and sometimes slightly higher than, that of the other FRO baselines, which is reasonable because the more conservative prediction intervals lead to a larger uncertainty set and a more computationally demanding downstream TSRO model. Therefore, the advantage of RTRO does not come from cheaper individual solves. Instead, the daily total solution time is reduced mainly because the number of re-optimizations is drastically decreased. This confirms that RTRO acts as a selective receding-horizon re-optimization mechanism.

Table~\ref{tab:online_stats} reports the multi-day statistical results over the typical-day set, the extreme-day set, and all test days. The same trend is observed across all subsets. CVaR-DFL-FRO achieves the best cost and $\mathrm{CVaR}_{90\%}$, while the proposed CVaR-DFL-RTRO remains very close to it and clearly outperforms the non-CVaR baselines. Over all test days, the performance gap between CVaR-DFL-RTRO and CVaR-DFL-FRO is below 0.5\% for both cost and $\mathrm{CVaR}_{90\%}$ on the two systems, whereas the total daily solution time is reduced by about 92\% on the IEEE 33-bus system and 88\% on the IEEE 69-bus system. Overall, the proposed method achieves near-FRO operational quality with only marginal performance sacrifice, while alleviating the majority of the online computational cost.

\subsection{Representative Dispatch Results}
\label{subsec:dispatch_results}

\begin{figure}[t]
\centering 
\includegraphics[width=8.4cm]{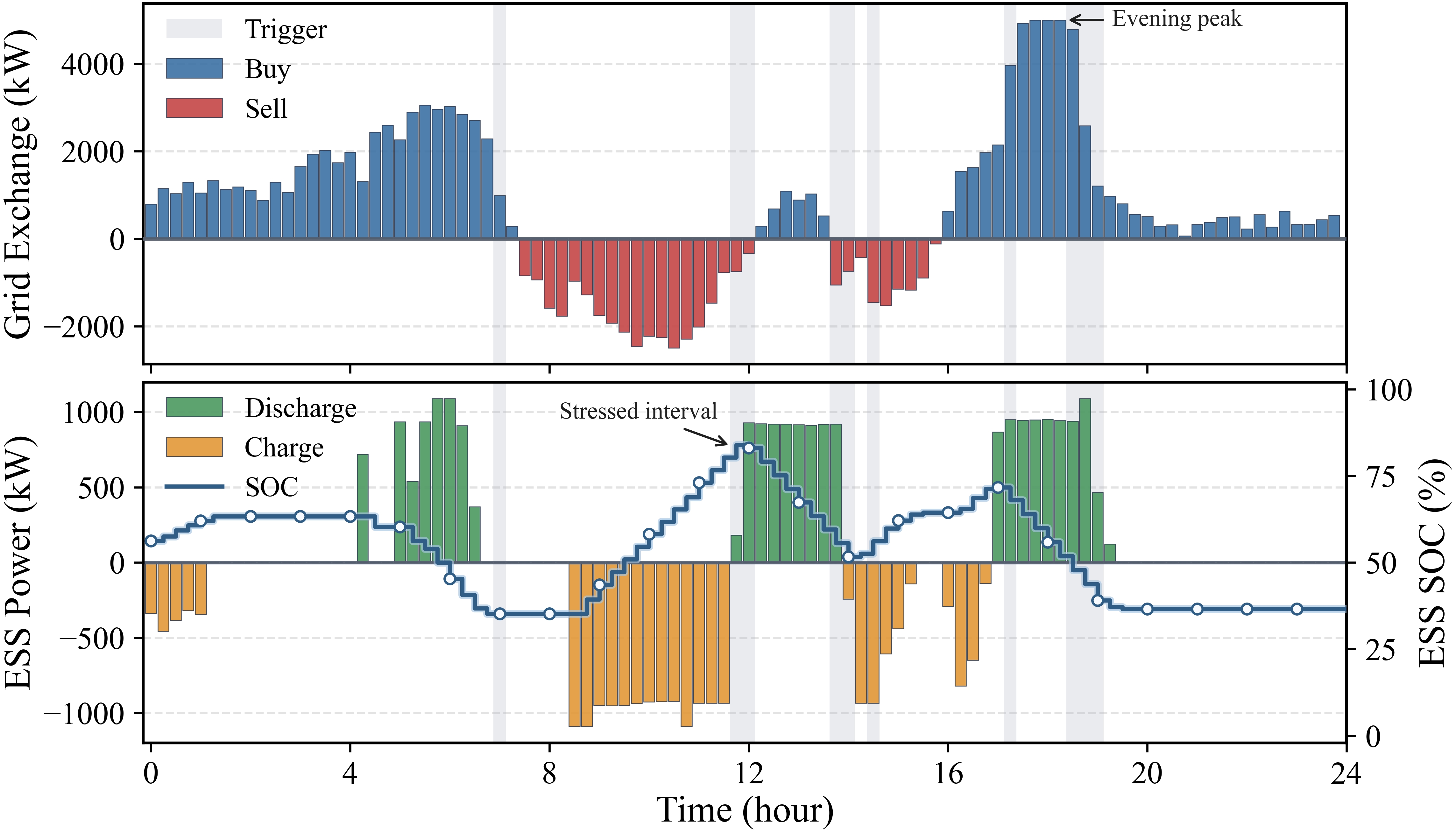}
\caption{24 h dispatch results of the proposed method on the representative extreme day.}
\label{fig:dispatch}
\vspace{-1.5mm}
\end{figure}

Fig.~\ref{fig:dispatch} shows the 24 h dispatch results of the proposed CVaR-DFL-RTRO method on the representative extreme day. The upper panel presents the power exchange with the main grid, while the lower panel shows the ESS charging/discharging power and the state-of-charge (SOC) trajectory. The RTRO activations are concentrated around periods with rapid net-load changes and enlarged uncertainty, rather than being uniformly distributed over the day, which is consistent with the role of the risk trigger.

From the dispatch profile, the microgrid relies heavily on power purchases from the main grid during high-demand intervals, particularly around the evening peak, whereas power exports dominate during surplus-generation periods. The ESS does not merely perform routine charge--discharge shifting; instead, its behavior is consistent with a risk-aware hedging strategy. In particular, the SOC is raised before stressed intervals and then released during subsequent high-risk periods, which helps preserve operational flexibility under the downstream TSRO model. The SOC trajectory remains consistent with the charging and discharging actions and stays within the admissible range throughout the day.

Although Fig.~\ref{fig:dispatch} mainly highlights grid exchange and ESS coordination, DLC remains available as a second-stage recourse and provides additional flexibility during stressed periods.

\section{Conclusion}
\label{sec:conclusion}

This paper developed a CVaR-guided decision-focused framework for two-stage robust microgrid operation, in which probabilistic load forecasting, robust dispatch, and online schedule adaptation were coordinated. By constructing the uncertainty set directly from prediction intervals, the proposed method established a data-driven connection between load uncertainty characterization and the downstream TSRO model. To enable such a forecast-decision closed loop, a convex regularized surrogate TSRO model and a smooth regret-based loss were introduced so that operational feedback could be propagated to the forecasting model through KKT-based implicit differentiation. In addition, a risk-triggered re-optimization mechanism was developed to selectively update the schedule, thereby reducing unnecessary online re-optimization.

Case studies on the modified IEEE 33-bus and 69-bus microgrids showed that the proposed method improved both forecasting quality and downstream decision quality. Compared with full re-optimization, the proposed RTRO mechanism kept the cost and CVaR$_{90\%}$ degradation below 0.5\% while reducing the daily total solution time by about 92\% on the IEEE 33-bus system and 88\% on the IEEE 69-bus system. These results indicate that probabilistic load forecasting and two-stage robust decision-making should be designed in a coordinated rather than sequential manner. The proposed framework provides a practical route for unifying probabilistic forecasting, two-stage robust dispatch, and selective online re-optimization within a decision-focused microgrid operation framework.

\bibliographystyle{IEEEtran}
\bibliography{reference}

\end{document}